# Overcoming the diffraction limit on the size of dielectric resonators using an amplifying medium

I. V. Doronin,[1,2] E. S. Andrianov[1,2], A. A. Zyablovsky[1,2*]

[1]Dukhov Research Institute of Automatics (VNIIA), 22 Sushchevskaya, Moscow, Russia, 127055

[2]Moscow Institute of Physics and Technology, 9 Institutskiy pereulok, Moscow, Russia, 141700

*zyablovskiy@mail.ru

Existing methods for the localization of light at the nanoscale use either a structure with negative permittivity, by exploiting subwavelength plasmonic resonances, or a dielectric structure with a high refractive index, which reduces the wavelength. In this paper, we provide an alternative to these two methods based on a modification of the modes of dielectric resonators by means of an active medium. We show that an active medium can promote subwavelength light localization in the dielectric structure. We consider a dielectric layer of size substantially smaller than a half-wavelength of light in the dielectric medium, and demonstrate that at a certain value of gain in the active medium, the phase change on reflection at the layer boundaries compensates for the change in phase due to propagation over the layer. At this value of the gain, the gain-assisted mode forms, in which the phase shift during a round trip of the electromagnetic wave is zero. This gain-assisted mode exists only at a positive gain in the dielectric medium, and can be used to create dielectric lasers and sensors of subwavelength size.

**Introduction**

Subwavelength light localization is of great importance from the points of view of both the possible applications and fundamental physics. Among the practical applications are high-sensitivity sensors (*1-4*), biochemical microscopy and sensing (*5-11*), photonics and optoelectronics (*12-20*). Biochemical sensing and microscopy take advantage of localization since it enables unprecedented resolution, whereas photonic devices benefit from miniaturization because it facilitates a decrease in the response time. The common theme in these applications is focusing light within a small volume, either for high-precision measurements or fast response times. There is also interest in the theoretical investigation and experimental creation of strong (*21-33*) and ultra-strong light-matter interactions (*34-36*).

The use of dielectric structures for subwavelength localization has been deemed impossible, since the Helmholz equation restricts the minimal possible size of a cavity based on a dielectric structure placed in a vacuum. If the dielectric permittivity $\varepsilon$ of the cavity material is real, the cavity based on a dielectric slab needs to be at least as large as $l_0 = \lambda / 2\sqrt{\varepsilon}$ (*37*). In this case, the phase acquired by the light after a full round trip over the cavity is equal to $2\pi$. Similar restrictions apply to other types of dielectric resonator, for example dielectric Mie particles (*38*). Using materials with high values of dielectric permittivity enables the cavity size to be reduced (*38-43*). However, although the creation of subwavelength dielectric structures made of high-index dielectric materials is a significant step towards optical field localization from a practical point of view (*38-40, 44-48*), it provides a quantitative rather than a qualitative solution to the problems arising due to the diffraction limit.

The subwavelength localization of light can be achieved via surface plasmon resonances (*1, 2, 8-12, 19, 20, 49-54*). This type of resonance only arises in structures containing materials with both positive and negative permittivity. This enables eigenmodes in which the amplitude of the electric field reaches a maximum near the interface between two materials with permittivities of opposite signs and decreases with the distance from this interface (*53*). At appropriate values of the dielectric permittivity, this decay causes subwavelength localization of light. However, plasmonic nanostructures typically have large ohmic losses (*55, 56*), which gives rise to the necessity of using gain materials for nanoscale light localization.

Until recently the gain media used in such structures have been considered as materials that only reduce the losses in the electromagnetic (EM) modes that are formed by the dielectric/plasmonic structure itself (*56-58*). It has only recently been understood that gain materials can substantially change the eigenmodes of the optical system (*59, 60*). For example, the use of amplifying and absorbing layers allows for the creation of an optical parity-time (PT) symmetrical structure (*61-75*), in which the change in the gain coefficient can lead to a spontaneous symmetry breaking of the eigenstates (*62-65*). The use of these structures makes it possible to create new types of lasers (*26, 33, 60, 76-80*), sensors (*81-84*), and waveguide systems (*85-88*).

Inspired by these achievements, the question arises as to whether it is possible to achieve subwavelength light localization in the dielectric structure by using an active medium. In this paper, we give a positive answer to this question.

We demonstrate that the diffraction limit can be overcome in a dielectric structure with an active medium. We consider a dielectric layer of subwavelength size containing an active medium, and demonstrate that in this system, a gain-assisted (GA) mode forms when the gain exceeds a certain value. In this mode, the phase shift arising due to the propagation of light through the layer is compensated by the negative phase shift on reflection at the boundaries of the active medium, meaning that the mode has zero round trip phase. The GA mode exists even in the limit of infinitely small layer thickness, and is manifested in the transmission and reflection spectra of the dielectric layer. With sufficient gain, lasing can occur in the GA mode in the subwavelength dielectric layer. These results pave the way for the creation of a new generation of subwavelength-sized lasers and sensors.

**Scattering matrix for a dielectric layer with active medium**

We study light propagation over a dielectric layer of thickness $l$ placed in free space. We consider that this layer contains an active medium of two-level atoms. In this case, the magnetic permeability of the layer $\mu = 1$, and the dielectric permittivity of the layer $\varepsilon$ is given by the Lorentz expression (*89, 90*):

$$\varepsilon(\omega) = \varepsilon_0 + \frac{\alpha}{\omega - \omega_0 + i\gamma} . \qquad (1)$$

The dielectric permittivity in (1) can be derived from the Maxwell-Bloch equations (*89*) for the electromagnetic field and the medium of two-level atoms, and obeys the Kramers-Kronig relations (*89, 91*). Here, $\varepsilon_0$ is the dielectric permittivity of the matrix ($\varepsilon_0 > 0$), and $\omega_0$ and $\gamma$ are the transition frequency and the linewidth of the two-level atoms, respectively. $\alpha = 4\pi |\mathbf{d}|^2 n_c D_0 / \hbar$ is obtained from Maxwell-Bloch equations (*89*), where $n_c$ is a concentration of two-level atoms, $\mathbf{d}$ is the dipole moment of the atomic transition, and $D_0$ is the population inversion of the two-level atoms (*89*). At $D_0 > 0$, the population of the excited state is larger than the population of the ground state. In this case, $\text{Im}\,\varepsilon(\omega_0) < 0$ ($\alpha > 0$), and the medium amplifies the EM wave propagating through it. At $D_0 < 0$, the medium absorbs EM waves.

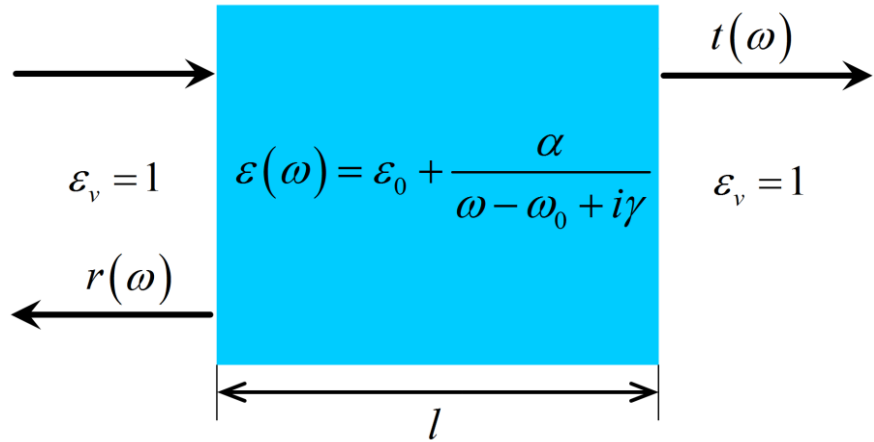

Figure 1. Scheme of the system under consideration.

We consider the scattering of EM waves by the layer (Fig. 1). The scattering matrix for this system is (*92*):

$$S(\omega) = \begin{pmatrix} r_L(\omega) & t(\omega) \\ t(\omega) & r_R(\omega) \end{pmatrix} \qquad (2)$$

where $t(\omega)$ is the transmission coefficient, and $r_L(\omega)$ and $r_R(\omega)$ are the reflection coefficients from the left and right sides of the layer, respectively. Due to the mirror symmetry of the system, these coefficients are equal to each other, $r_L(\omega) = r_R(\omega) = r(\omega)$. The transmission coefficients in the two opposite directions are equal to each other as well, in accordance with the Lorentz reciprocity theorem (*93*). In this system, the transmission and reflection coefficients are (*89*):

$$t(\omega) = \frac{t_{1\varepsilon} t_{\varepsilon 1} e^{i\frac{\omega}{c}\sqrt{\varepsilon(\omega)}\,l}}{1 - r_{\varepsilon 1}^2 e^{2i\frac{\omega}{c}\sqrt{\varepsilon(\omega)}\,l}}, \qquad (3a)$$

$$r(\omega) = r_{1\varepsilon} + \frac{r_{\varepsilon 1} t_{1\varepsilon} t_{\varepsilon 1} e^{2i\frac{\omega}{c}\sqrt{\varepsilon(\omega)}\,l}}{1 - r_{\varepsilon 1}^2 e^{2i\frac{\omega}{c}\sqrt{\varepsilon(\omega)}\,l}}. \qquad (3b)$$

Here, $r_{1\varepsilon} = \dfrac{1-\sqrt{\varepsilon(\omega)}}{1+\sqrt{\varepsilon(\omega)}}$ and $r_{\varepsilon 1} = \dfrac{\sqrt{\varepsilon(\omega)}-1}{\sqrt{\varepsilon(\omega)}+1}$ are the coefficients of reflection at the vacuum-dielectric and dielectric-vacuum boundaries, respectively (*37*). $t_{1\varepsilon} = \dfrac{2}{\sqrt{\varepsilon(\omega)}+1}$ and $t_{\varepsilon 1} = \dfrac{2\sqrt{\varepsilon(\omega)}}{\sqrt{\varepsilon(\omega)}+1}$ are the coefficients of transmission through the vacuum-dielectric and dielectric-vacuum boundaries, respectively (*37*).

The scattering matrix $S(\omega)$ has poles at certain frequencies determined by the condition (*89*):

$$\left(\frac{\sqrt{\varepsilon(\omega)}-1}{\sqrt{\varepsilon(\omega)}+1}\right)^2 \exp\left(2i\frac{\omega}{c}\sqrt{\varepsilon(\omega)}\,l\right) = 1. \qquad (4)$$

For an absorbing layer ($\alpha < 0$), the frequencies of all poles are in the lower half of the complex frequency plane. An increase in the population inversion in the active medium (i.e., an increase in $\alpha$) causes the poles to move upward in the complex frequency plane. When at least one of the poles lies in the upper half of the complex frequency plane, lasing takes place in the system (*76, 89, 92, 94*).

The condition (4) is reduced to amplitude and phase conditions (*89, 95, 96*). The amplitude condition is obtained by taking the modulus of the both sides of (4) (*89*) and requires that the amplification must overcome the loss (*89, 95, 96*). The phase condition requires that the phase shift of the EM field per one round trip over the layer is a multiple of $2\pi$ (*89, 95, 96*), i.e.

$$2\frac{\omega}{c}\operatorname{Re}\sqrt{\varepsilon(\omega)}\,l + 2\arg\left(\frac{\sqrt{\varepsilon(\omega)}-1}{\sqrt{\varepsilon(\omega)}+1}\right) = 2\pi m \qquad (5)$$

The existence of a Fabry-Perot mode in the layer is associated with fulfillment of the phase condition (5). The frequency of the mode is determined by the condition (5). At mode frequencies determined by (5), the transmission coefficient $t(\omega)$ (Eq. (3a)) reaches one, while the reflection coefficient $r(\omega)$ (Eq. (3b)) reaches zero (see Fig. 2). It is usually assumed that the minimal thickness of a layer that supports a Fabry-Perot mode is $l_0 = \lambda/(2\operatorname{Re}\sqrt{\varepsilon})$. Therefore, the minimal size of the dielectric resonator appears to be limited by the diffraction.

However, the resonant phase condition (5) is formally satisfied for $m=0$. In this case, the phase shift of the EM field for each round trip over the layer is zero. For a passive layer ($\alpha = 0$), this situation corresponds to zero layer thickness $l=0$ (see Fig. 2). In this limit, the equalities $t(\omega)=1$ and $r(\omega)=0$ not only hold true, but are fulfilled for all frequencies $\omega$. This behavior is associated with the propagation of light through empty space, and thus the resonance for $m=0$ is usually disregarded.

In the following section, we show for the first time that in the presence of the positive imaginary part of dielectric permittivity, resonance for $m=0$ becomes possible for a layer of finite thickness.

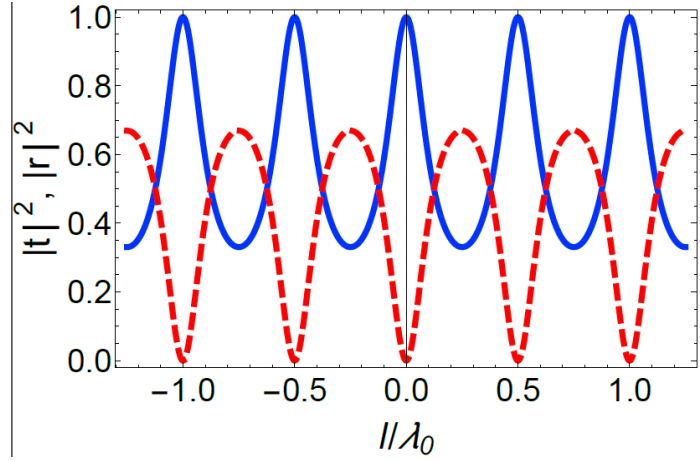

Figure 2. Dependencies of transmission (solid blue line) and reflection (dashed red line) coefficients of a dielectric layer on the thickness of the layer.

**Condition for the existence of a gain-assisted mode in the subwavelength layer**

The phase condition (5) for $m=0$ means that the total phase acquired during one round trip inside the cavity is equal to zero. This total phase shift $\Delta\phi$ is equal to the sum of the phase shifts arising from propagation, $\Delta\phi_{pr} = 2l\,\text{Re}\left(\sqrt{\varepsilon(\omega)}\right)\omega/c = 4\pi l\,\text{Re}\left(\sqrt{\varepsilon(\omega)}\right)/\lambda$, and from reflection, $\Delta\phi_{ref}(\omega) = 2\arg\left(\dfrac{\sqrt{\varepsilon(\omega)}-1}{\sqrt{\varepsilon(\omega)}+1}\right)$.

In a *finite* layer ($l>0$) with *real* permittivity, the zero phase shift condition is not satisfied. Indeed, for a dielectric layer with $\varepsilon > 1$ placed in a vacuum with $\varepsilon_v = 1$, we have $\Delta\phi_{pr} = 2l\sqrt{\varepsilon(\omega)}\,\omega/c = 4\pi l\sqrt{\varepsilon(\omega)}/\lambda > 0$, whereas $\Delta\phi_{ref} = 0$. Hence, to satisfy the resonance condition, $\Delta\phi_{pr}$ must be at least as large as $2\pi$. In other words, the size of the dielectric layer needs to be at least half of the wavelength of light in the medium.

This situation changes significantly if the permittivity has a positive imaginary part. The dependence of the phase shift on reflection, $\Delta\phi_{ref}$, on $\alpha$ at the resonance frequency ($\omega = \omega_0$) is shown in Fig. 3a. It can be seen that the phase shift $\Delta\phi_{ref}$ (red dashed line in Fig. 3a) is negative for $\alpha > 0$, i.e., for an amplifying medium. The phase shift arising during propagation, $\Delta\phi_{pr}$, remains positive for all $\alpha$ (blue solid line in Fig. 3a) because it depends only on the real part of the wave vector, $k(\omega) = \omega\sqrt{\varepsilon(\omega)}/c$. As a result, using an active medium with a suitable value of $\alpha$ makes it possible to compensate for the phase shift arising from propagation by the phase shift on reflection. In this case, the total phase shift (black dotted line in Fig. 3a) arising during one round trip inside the cavity can be equal to zero. This enables resonance in a subwavelength dielectric layer of non-zero thickness. We call this resonance the gain-assisted (GA) mode.

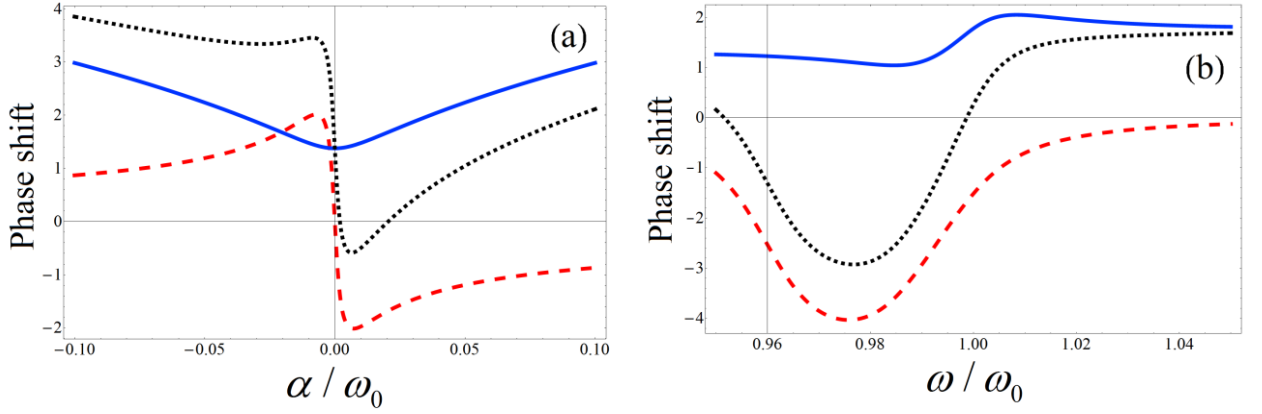

Figure 3. (a) Dependence of the phase shifts $\Delta\phi_{pr}$ (solid blue line), $\Delta\phi_{ref}$ (dashed red line) and $\Delta\phi_{tot} = \Delta\phi_{pr} + \Delta\phi_{ref}$ (dotted black line) on $\alpha$ at the transition frequency $\omega_0$ for $\varepsilon_0 = 1.2$ and $l = 0.1\lambda$. (b) Dependence of the phase shifts $\Delta\phi_{pr}$ (solid blue line), $\Delta\phi_{ref}$ (dashed red line), and $\Delta\phi_{tot} = \Delta\phi_{pr} + \Delta\phi_{ref}$ (dotted black line) on the frequency in the case of an amplifying medium, $\alpha = 0.02\omega_0 > 0$. The parameters are $l = 0.1\lambda_0$; $\gamma = 0.01\omega_0$; $\varepsilon_0 = 1.2$.

Using the Lorentz dependence for dielectric permittivity (1), we can calculate the value of $\alpha$ at which $\Delta\phi_{tot} = \Delta\phi_{pr} + \Delta\phi_{ref}$ is zero for a given layer thicknesses, i.e., we can find the threshold for the GA mode formation, $\alpha_{th\_m}$. As can be seen from Fig. 4a (solid blue line), this mode can form even when the layer thickness is much smaller than one half-wavelength. The frequency dispersion in the permittivity of the active medium (see Eq. (1)) ensures that at $\alpha > \alpha_{th\_m}$ the phase condition (5) is always fulfilled at some frequency near the transition frequency (see dotted black lines in Fig. 3b).

Notably, when the layer thickness exceeds one half-wavelength in the dielectric, the first Fabry-Perot mode ($\Delta\phi_{tot} = 2\pi$) appears in the layer (solid blue line in Fig. 4a). This mode exists for both negative and positive $\alpha$. That is, in contrast to the GA mode, the gain is not necessary for the existence of the Fabry-Perot mode. For $\alpha > \alpha_{th\_m}$ and when the layer thickness exceeds one half-wavelength in the dielectric, both the GA mode and the conventional Fabry-Perot mode exist in the layer (Fig. 4b). Note that the frequency of the GA mode (dashed red line in Fig. 4a) barely depends on the layer thickness, and is primarily determined by the transition frequency of the amplifying medium. In contrast, the frequencies of the conventional Fabry-Perot modes strongly depend on the layer thickness (solid red line in Fig. 4a).

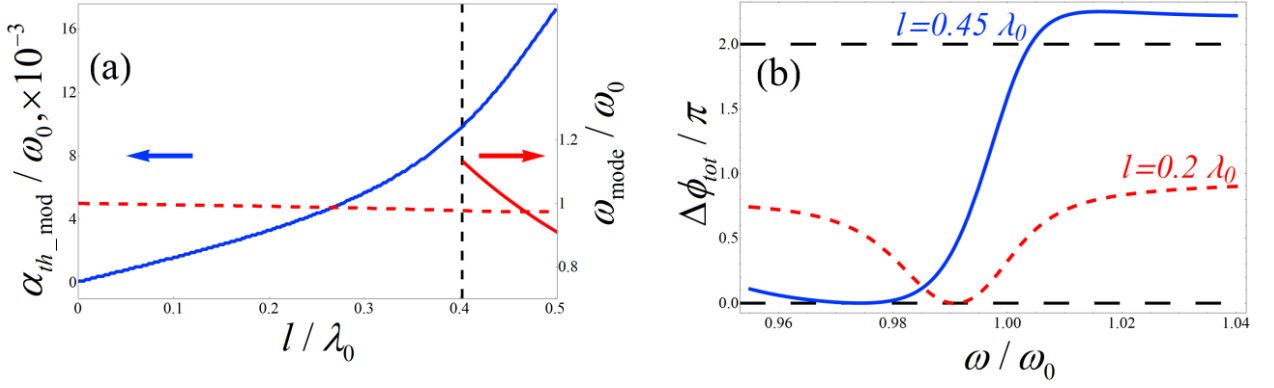

Figure 4. (a) Dependence of the threshold of GA mode formation ($\alpha_{th\_m}$) in a dielectric layer of thickness $l$. The solid blue line shows the threshold value of $\alpha_{th\_m}$, and the dashed red line is the mode frequency at $\alpha = \alpha_{th\_m}$. When the layer thickness exceeds one half-wavelength in the dielectric, the Fabry-Perot mode exists in the layer without the amplifying medium. The vertical dashed black line shows the layer thickness above which the Fabry-Perot mode exists in the passive layer ($\alpha = 0.0$). The solid red line shows the Fabry-Perot mode frequency for $\alpha = 0.0$. (b) Dependence of the total phase shift arising during one round trip inside a layer of thickness $l$ on the frequency. The solid blue line shows the dependence for $l = 0.45\lambda_0$; $\alpha = 0.013$. The dashed red line shows the dependence for $l = 0.2\lambda_0$; $\alpha = 0.00326$. $\gamma = 0.01\omega_0$, and $\varepsilon_0 = 1.2$. The horizontal dashed lines show the phase conditions for the GA mode ($\Delta\phi_{tot} = 0$) and the Fabry-Perot mode ($\Delta\phi_{tot} = 2\pi$).

### Lasing in the GA mode

In the previous section, we showed that the use of the active medium enables the GA mode in the subwavelength dielectric layer as long as the phase condition (5) is fulfilled for $m = 0$. However, the fulfillment of the phase condition (5) is not sufficient for lasing, which occurs when the frequency of at least one of the poles of the scattering matrix lies in the upper half of the complex frequency plane (*76, 89, 92, 94*).

Using Eq. (4), we find the poles of the scattering matrix (Fig. 5) and the lasing threshold for different layer thicknesses (Fig. 6). The lasing threshold $\alpha_{th\_las}$ is determined as the minimal value of the imaginary part of the dielectric permittivity at which the frequency of one of the scattering matrix poles becomes real (*76, 89, 92*). Our calculations show that depending on the layer thickness, lasing occurs either in the GA mode or in the conventional Fabry-Perot modes. When the layer thickness exceeds one half-wavelength in the dielectric, the pole of the scattering matrix corresponding to the Fabry-Perot mode has a smaller threshold value for $\alpha$ than the pole corresponding to the GA mode (Fig. 6c). In this case, lasing starts in the conventional Fabry-Perot mode. However, when the layer thickness is less than one half-wavelength in the dielectric, the lasing threshold for the GA mode is smaller than that of the Fabry-Perot modes, and lasing occurs in the GA mode.

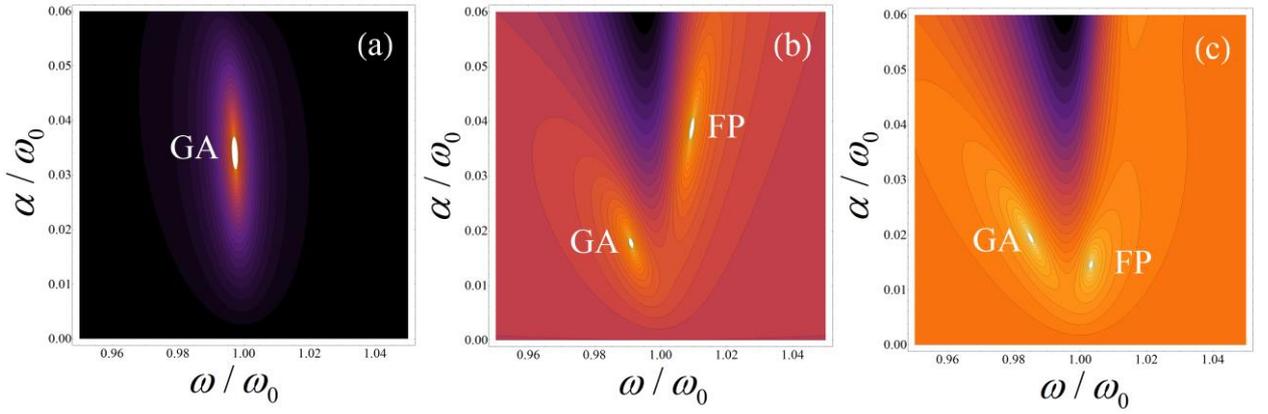

Figure 5. Dependences of the decimal logarithm of the absolute value of the transmission coefficient in (3a) on $\omega$ and $\alpha$. The layer thicknesses are (a) $l = 0.1\lambda_0$; (b) $l = 0.3\lambda_0$; and (c) $l = 0.45\lambda_0$. The letters GA and FP are used to mark the poles of the transmission coefficient corresponding to the GA mode and the Fabry-Perot mode, respectively. The reflection coefficient and the scattering matrix have poles at the same values of $\omega$ and $\alpha$; $\varepsilon_0 = 1.2$.

Lasing in the GA mode is possible even when the thickness of the dielectric layer with the active medium is much smaller than one half-wavelength (see Fig. 6). The lasing threshold in this case is inversely proportional to the layer thickness (gray line in Fig. 6).

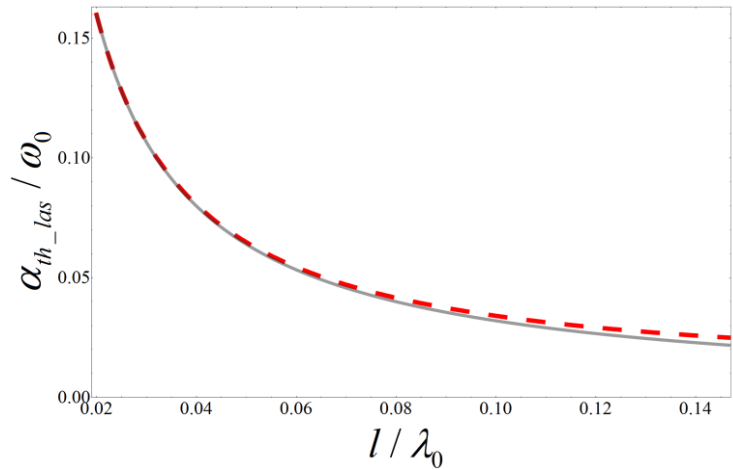

Figure 6. Dependence of the lasing threshold ($\mathrm{Im}\,\varepsilon(\omega_0) \Box \alpha$) on the layer thickness $l$ (dashed red line). The gray line shows fitting of the lasing threshold dependence by $l^{-1}$. $\varepsilon_0 = 1.2$. $\gamma = 0.01\omega_0$.

Lasing in the GA mode can take place in a dielectric layer with thickness $l \ll \lambda_0 / 2|n(\omega_0)|$, where $|n(\omega_0)| = \left|\sqrt{\varepsilon(\omega_0)}\right|$ is the absolute value of the refractive index, and $\lambda_0 = 2\pi c/\omega_0$ is the wavelength in free space at the transition frequency $\omega_0$. To illustrate this statement, we find the dependence of the layer thickness on the absolute value of the refractive index at which lasing starts. To do this, we recalculate the dependence in Fig. 6 using the expression (1) for dielectric permittivity. The result is shown in Fig. 7.

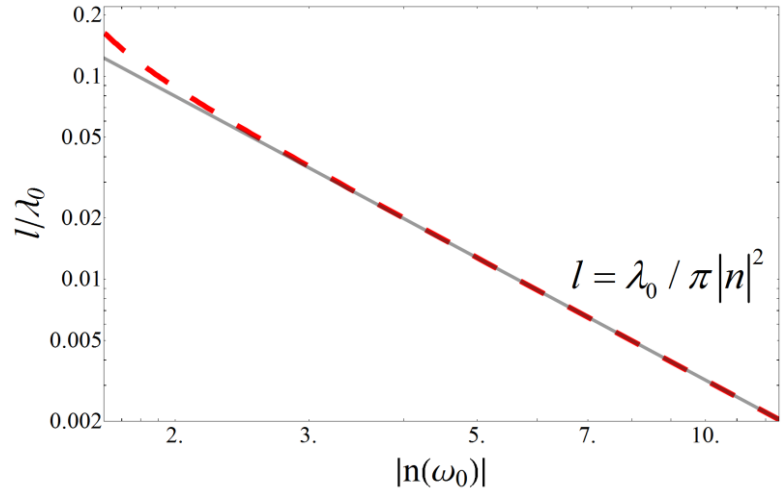

Figure 7. Dependence of the layer thickness $l$ on the absolute value of the refractive index $|n(\omega_0)| = |\sqrt{\varepsilon(\omega_0)}|$ at which lasing starts (dashed red line). The solid gray line is $l = \lambda_0 / \pi |n|^2$.

In the case of the conventional Fabry-Perot modes, the phase condition at the transition frequency $\omega_0$ is fulfilled when $l = m\lambda_0 / (2\sqrt{\varepsilon(\omega_0)})$, where $m = 1,...$. It is generally considered that the gain shifts the resonance frequency only slightly. In this case, to create a dielectric laser of thickness $l < \lambda_0 / 2$, it is necessary to use a material with $|n(\omega_0)| \approx 2l/\lambda_0$. That is, the thickness of the dielectric laser scales as $1/|n(\omega_0)|$ (solid gray line in Fig. 7). For the GA mode, the dependence of the layer thickness on the absolute value of the refractive index is sharper (dashed red line in Fig. 7) and at $|n(\omega_0)| \gg 1$ it can be approximated by $l = \lambda_0 / \pi |n|^2$ (solid gray line in Fig. 7). Hence, lasing in the GA mode can take place in a dielectric layer of thickness $l \ll \lambda_0 / 2|n(\omega_0)|$.

We can therefore conclude that an active dielectric layer much smaller than one half-wavelength can serve as a laser resonator.

**Observation of the GA mode and lasing in this mode**

In this section, we determine the experimental parameters necessary for lasing in the GA mode. To this end, we calculate the gain corresponding to the lasing threshold using the following expression (*97*):

$$G = -2\frac{\omega}{c} \text{Im} \sqrt{\varepsilon}. \qquad (6)$$

The threshold gain is inversely proportional to the layer thickness (Fig. 6). When the layer thickness is greater than $0.3\lambda_0$, the threshold gain takes experimentally achievable values ($\Box\, 2\times10^5\ cm^{-1}$) (*95*). This layer thickness is smaller than one half-wavelength, which indicates the possibility of overcoming the diffraction limit using currently existing amplifying media.

Future advances in amplifying media may give rise to active media with larger gain values, and thus further improve the localization effect.

Although lasing in a subwavelength dielectric layer requires exceptionally high gain coefficients, the signature of the GA mode can be detected experimentally even in media with a lower value of the gain coefficient. This is because the transmission and reflection spectra exhibit resonance behavior at frequencies corresponding to the poles of the scattering matrix, even if these poles have negative imaginary parts (*89, 92*). In this way, we can measure the transmission and reflection spectra for a subwavelength dielectric layer with an amplifying medium. These spectra demonstrate resonance behavior (see Fig. 8) near the frequency corresponding to the pole of the scattering matrix, which is achievable for a higher value of the gain (see Fig. 5). Resonance behavior can be clearly observed at a noticeably lower gain than that required for lasing. This makes it possible to detect the GA mode in experiments that are simpler than lasing experiments.

Note that the frequencies of the peaks in the transmission and reflection coefficients do not coincide with the transition frequency in the active medium. This is a manifestation of the reflection at the boundaries within a layer with complex dielectric permittivity, which leads to the formation of a pole corresponding to the GA mode in the transmission and reflection coefficients at a higher value of the gain. Fig. 8 shows the dependencies of the reflection and transmission coefficients on the frequency in the case of a dielectric layer of subwavelength size, $l = 0.2\lambda_0$. It can clearly be seen that the maxima for the reflection and transmission coefficients manifests at gain values lower than the lasing threshold gain of the GA mode.

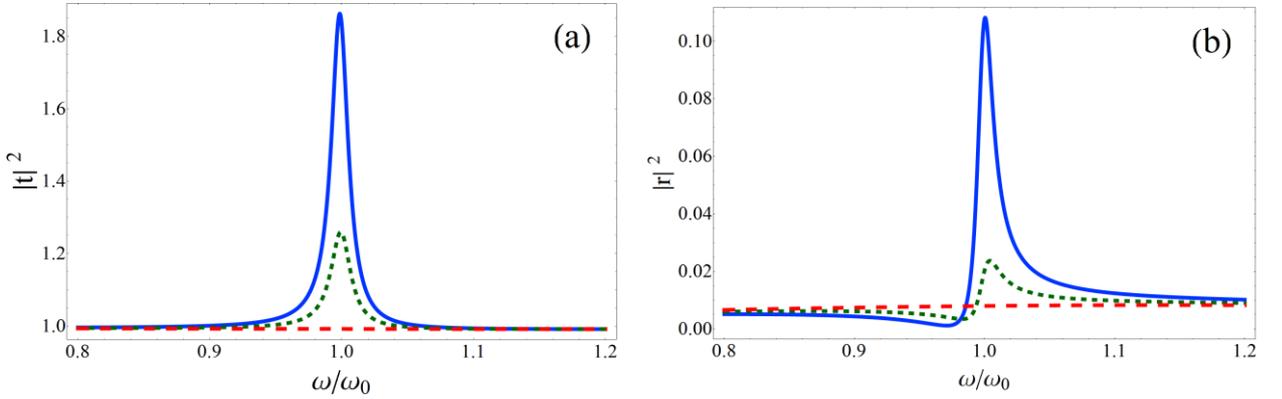

Figure 8. Dependences of the coefficients of (a) transmission and (b) reflection on the frequency for $l = 0.2\lambda_0$; $\varepsilon_0 = 1.2$. $\alpha = 0$ (dashed red lines); $\alpha = 0.002$ (dotted green lines); $\alpha = 0.005$ (solid blue lines). These values of $\alpha$ approximately correspond to gain values of $G = 0\,cm^{-1}$, $G = 2.2 \times 10^4\,cm^{-1}$, and $G = 5.6 \times 10^4\,cm^{-1}$, respectively.

**Conclusion**

We have established a new type of localized mode in a dielectric structure with an amplifying medium that makes it possible to overcome the diffraction limit on the size of dielectric structures. We consider a dielectric layer with a size that is substantially smaller than the half-wavelength of light in the dielectric medium, and demonstrate that the use of an amplifying medium enables the formation of a gain-assisted mode that is characterized by a zero round trip phase for the EM wave. This is achieved because the phase change on reflection at the boundaries of the active medium and the vacuum compensates for the change in phase due to propagation over the layer. We demonstrate that there is a gain above which gain-assisted mode lasing in the subwavelength dielectric layer takes place. The lasing frequency is about equal to the transition frequency in the active medium, $\omega_0$. The thickness of the layer $l$ supporting gain-assisted mode lasing is about $\lambda_0/\pi |n(\omega_0)|^2$ in contrast to the dependence $l \Box \lambda_0/2|n(\omega_0)|$ for Fabry-Perot mode. That is, the use of the amplifying media enables us to achieve lasing with stronger localization of the EM field in terms of $|n(\omega_0)|$. The results presented here open the way for a new generation of subwavelength-sized dielectric lasers and sensors.

**Funding.** The study was supported by a Grant from Russian Science Foundation (Project no. 20-72-10057).

**Acknowledgments.** The study was financially supported by a Grant from Russian Science Foundation (Project No. 20-72-10057). I.V.D. and E.S.A. thank foundation for the advancement of theoretical physics and mathematics "Basis".